\documentclass[11pt]{article}
\pdfoutput=1 

\usepackage{jheppub} 

\usepackage{pict2e}
\usepackage{physics}
\usepackage{amsfonts}
\usepackage{mathtools}
\usepackage{amssymb,bbm}
\usepackage{dsfont}
\usepackage{subcaption}
\usepackage{braket}

\title{\boldmath Janus and  RG-interfaces in minimal 3d gauged supergravity}

\author[a]{Michael Gutperle,}
\author[a]{Charlie Hultgreen-Mena}

\affiliation[a]{
Mani L. Bhaumik Institute for Theoretical Physics, Department of Physics and Astronomy,\\ University of California, Los Angeles, CA 90095, USA
}

\emailAdd{gutperle@ucla.edu}
\emailAdd{charliemena25@g.ucla.edu}

\abstract{In this paper we find solutions of minimal $d=3,N=2$ gauged supergravity corresponding to Janus and RG-flow interfaces. We use  holography  to calculate symmetric and interface entanglement entropy as well as reflection coefficients and confirm that a recently proposed \cite{Karch:2024udk} inequality  involving these quantities is satisfied for the solutions found here.}

\begin{document} 
\maketitle
\flushbottom

\section{Introduction}
Janus solutions are solutions of supergravity theories which describe interface CFTs  in the AdS/CFT correspondence. The first such solution \cite{Bak:2003jk} was constructed in  type IIB supergravity describing $N=4$ Super Yang-Mills theory with the gauge coupling jumping across a planar interface. There are two different approaches to constructing such solutions, one is the top-down approach where ten or eleven dimensional solutions type II or M-theory are constructed as  products involving AdS and spherical factors warped over a Riemann surface with boundary (see e.g. \cite{DHoker:2007zhm,DHoker:2006qeo,DHoker:2007hhe,DHoker:2008lup,DHoker:2009lky}).  A guiding principle is to look for solutions preserving half the number of supersymmetries of the $AdS$ vacua which allows to construct the explicit solutions from  harmonic functions on the Riemann surface with certain boundary conditions.

A second approach is to construct supersymmetric Janus  solutions in lower dimensional gauged supergravities (see e.g. \cite{Clark:2005te,Bobev:2013yra,Suh:2011xc,Chiodaroli:2011nr,Gutperle:2017nwo,Pilch:2015dwa,Karndumri:2016tpf,Bobev:2019jbi,Bobev:2020fon,Karndumri:2024uxz,Karndumri:2024jib,Anabalon:2022fti,Karndumri:2021pva}). Such solutions are often easier to obtain since all fields only depend on a single AdS-slicing coordinate and the Killing spinor equations are simpler. In many cases, the gauged supergravity theories are consistent truncations of ten and eleven dimensional supergravities and lower dimensional solutions can be uplifted. In addition, the simpler form of the solution allows to calculate holographic observables and handle holographic renormalization more easily.

Another reason to consider lower dimensional gauged supergravity is that these theories often have more than one AdS vacuum, coming from multiple extrema of the scalar potential. Apart from the maximally supersymmetric vacuum, the other vacua can have a lower number or no supersymmetry and in general will correspond to an AdS space with different values of the  cosmological constant, which translates into CFTs with a different central charge. For an ansatz with a Poincare sliced metric, it is possible to construct solutions which correspond to holographic RG-flows relating the two CFTs (see e.g. \cite{Freedman:1999gp,Girardello:1999bd,Bianchi:2000sm,Bobev:2009ms}). Using the Janus AdS-
slicing it is possible to construct RG-flow interfaces which describe an interface between two different CFTs which are related by an RG-flow. On  the field theory side RG-flow interfaces were discussed in \cite{Brunner:2007ur,Gaiotto:2012np,Dimofte:2013lba,Poghosyan:2014jia,Konechny:2016eek,Giombi:2024qbm} and examples of holographic RG-flow 
interface solutions are \cite{Gutperle:2012hy,Chen:2021mtn,Gutperle:2022fma,Arav:2020asu}. The goal of this paper is to find Janus and RG-flow solutions in one of the  minimal theories  in three dimensions, namely $N=2,d=3$ gauged supergravity, in order to have a set of simple (numerical) solutions for which holographic observables can be calculated. The ones we focus on in this paper are the interface 
entropy $\ln(g_A)$ for an entangling surface which is symmetric about the interface, the effective central charge $c_{eff}$ associated with the entanglement entropy where the entangling 
surface ends at the interface and the reflection coefficient $c_{LR}$ for the scattering of stress tensor modes off the interface. We use the solutions to test bounds and relations between the latter two quantities which have been investigated recently \cite{Karch:2023evr,Karch:2024udk,Baig:2024hfc}. The structure of this note is as follows: 
In section \ref{sec2} we review the $N=2,d=3$ gauged supergravity for which we will construct Janus and RG-flow solutions. In section \ref{sec3}, we set up the equations of 
motion for an $AdS_2$ slicing ansatz and generate families of 
numerical solutions both for Janus solutions which have the same CFT on both sides of the interface and RG-flow interfaces between two different CFTs. In section \ref{sec4} we briefly 
review the holographic observables we calculate and plot the results for some example solutions. While the results for the minimal   $N=2,d=3$ gauged supergravity are numerical, there exists a solution of $N=8,d=3$ gauged supergravity found
previously by one of the authors in \cite{Chen:2020efh} which is exact and preserves half the supersymmetries. In section \ref{sec5}  we calculate the holographic observables and observe that the relation between $c_{eff}$ and $c_{LR}$, which 
was pointed out to hold for the ten dimensional supersymmetric Janus solutions in  \cite{Baig:2024hfc} also hold for the  solutions constructed in this paper. We close the note with a discussion of the results and some future research directions in section \ref{sec6}.

\section{$N=2, d=3$ gauged supergravity}
\label{sec2}

In this note we will use a minimal form of  $N=2,d=3$ gauged supergravity where the bosonic sector is given  by three dimensional gravity, a complex scalar and a Chern-Simons $U(1)$ gauge field.  We will set the fermionic degrees of freedom to vanish and use the fermionic supersymmetry variations to test whether supersymmetries are preserved by the solutions.
The action was constructed in \cite{Deger:1999st}  and we will follow the conventions of \cite{Deger:2002hv,Arkhipova:2024iem}.
The Lagrangian is given by
\begin{align}
    S= \frac{1}{ 4 } \int d^3 x \sqrt{g} \Big( R- \frac{4|D_\mu  \Phi|^2 }{  a^2 (1- |\Phi|^2)^2} -  V(\Phi) \Big) + \frac{1}{ 4 m a^4} \int A\wedge dA
\end{align}
The covariant derivative coupling the complex scalar and the $U(1)$ gauge field is given by
\begin{align}
    D_\mu \Phi = \partial_\mu \Phi + i A_\mu \Phi
\end{align}
The scalar potential can be most conveniently expressed using the following parameterization
\begin{align}
    C= \frac{1+|\Phi|^2}{ 1-|\Phi|^2}, \quad  S= \frac{2 \Phi }{ 1-|\Phi|^2}
\end{align}
and is given by
\begin{align}
    V= 8 m^2 C^2(2 a^2 |S|^2 -C^2)
\end{align}
The Chern-Simons gauge field couples to the phase of the complex scalar field $\Phi$
\begin{align}
    \Phi = |\Phi| e^{i \theta}
\end{align}
It is convenient to introduce one more change of variable for the absolute value of the scalar field
\begin{align}
    |\Phi|= \tanh \left(\frac{a\phi}{2\sqrt{2}}\right)
\end{align}
which implies
\begin{align}
    C = \cosh \left(\frac{a \phi}{ \sqrt{2}}\right), \quad |S| =\sinh \left(\frac{a \phi}{ \sqrt{2}}\right)
\end{align}
The action can then be written in terms of the fields $\phi, \theta$
\begin{align}\label{actiona}
S&= \frac{1}{4} \int d^3 x \sqrt{g} \Big( R- \frac{1}{ 2} \partial_\mu \phi \partial^\mu \phi -  V(\phi)\Big)  \nonumber \\
& \;\;+  \frac{1}{4} \int d^3 x \sqrt{g} \Big( -\frac{\sinh^2 \phi}{ a^2}  (\partial_\mu \theta + A_\mu) (\partial^\mu \theta+A^\mu)+  \frac{1}{ 4 m a^4} \int A\wedge dA
\end{align}
In order to construct Janus and RG-flow interface solutions it is possible to set  $A_\mu=\theta=0$ consistently. The action is then given by the first line in (\ref{actiona}), i.e. three dimensional gravity minimally coupled to a real scalar field $\phi$ with a potential  $V$
\begin{align}
   V(\phi)=  -8m^2 \cosh^2\left(\frac{a\phi}{ \sqrt{2}}\right)\left[\cosh^2\left(\frac{a\phi}{ \sqrt{2}}\right)-2 a^2 \sinh^2 \left(\frac{a\phi}{ \sqrt{2}}\right) \right]
\end{align}
The $d=3,N=2$ supersymmetry transformation of the gravitino and dilatino for the truncated  Lagrangian takes the following form 
\begin{align}\label{susytransf}
    \delta \psi_\mu &=(\partial_\mu +\frac{1}{ 4} \omega_\mu^{ab}\gamma_{ab})\epsilon+\frac{1}{ 2} W\gamma_\mu \epsilon \nonumber\\
    \delta\lambda &=\frac{1}{ 2}(-\gamma^\mu \partial_\mu \phi -\frac{2}{a} \frac{\partial W}{ \partial \phi} )\epsilon
\end{align}
where  the superpotential is given by
\begin{align}
    W=2m \cosh^2 \left( \frac{a\phi}{ \sqrt{2}}\right)
\end{align}
and the potential is related to the superpotential by the following relation
\begin{align}
    V= 2\left(\frac{\partial W}{ \partial \phi}\right)^2 -2 W^2
\end{align}

\begin{figure}[ht]
     \centering
     \begin{subfigure}[b]{0.32\textwidth}
         \centering
    \includegraphics[width=\textwidth]{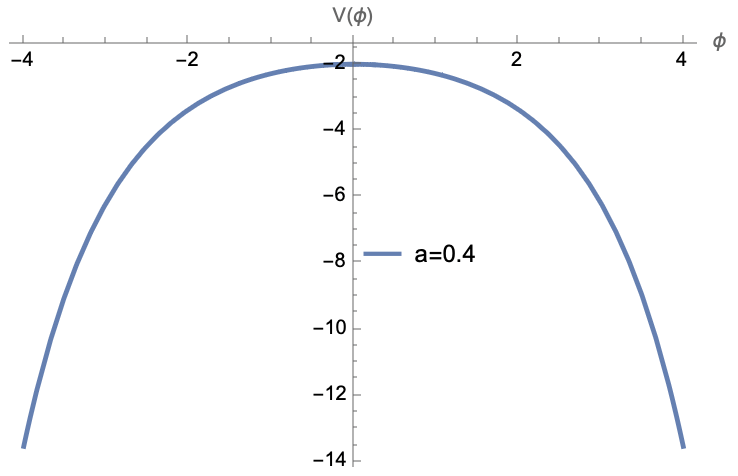}
         \caption{$V(\phi)$ for $a=0.4$}
         \label{1-a}
     \end{subfigure}
     \hfill
     \begin{subfigure}[b]{0.32\textwidth}
         \centering
         \includegraphics[width=\textwidth]{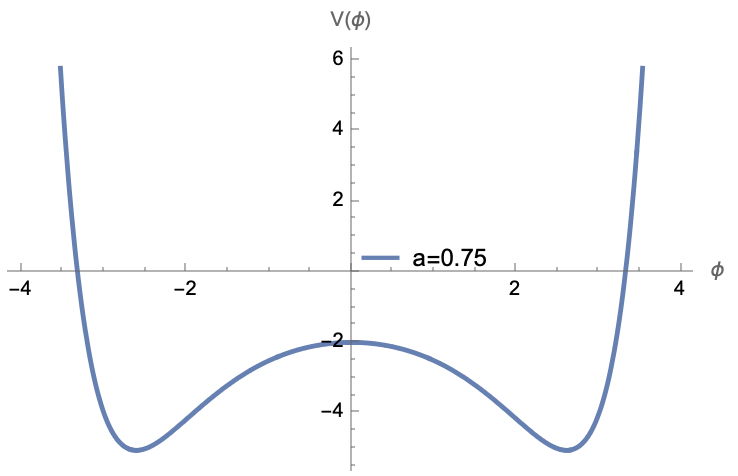}
         \caption{$V(\phi)$ for $a={0.75}$ }
     \end{subfigure}
     \hfill
\begin{subfigure}[b]{0.32\textwidth}
         \centering
         \includegraphics[width=\textwidth]{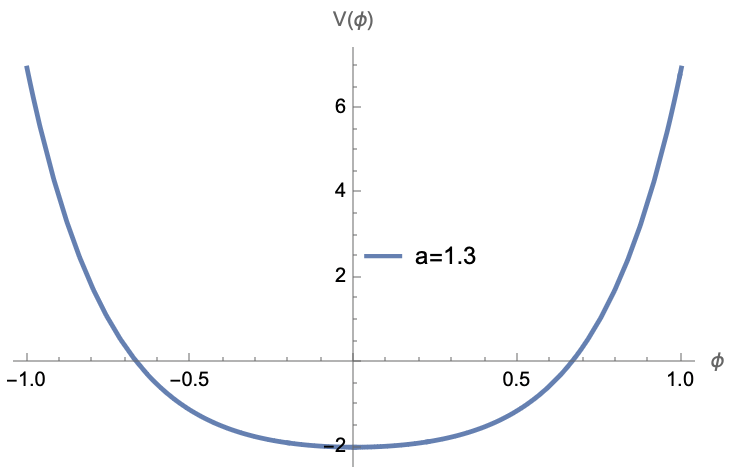}
         \caption{$V(\phi)$ for $a=1.3$  }
     \end{subfigure}
        \caption{Example of potential $V(\phi)$ for three cases (a) $a< \frac{1}{ \sqrt{2}}$, (b) $\frac{1}{ \sqrt{2}}<a<1$, (c) $a>1$}
        \label{figure:plotV}
\end{figure}

Note that $m$  only appears as an overall multiplicative factor in the potential, we will set $m=\frac{1}{2}$ which leads to a unit radius $AdS_3$ vacuum for $\phi=0$.  The shape of the potential and the number  and nature  of extrema depend on the parameter $a$, representative plots for the three different cases are shown in figure \ref{figure:plotV}.

\begin{figure}[ht]
     \centering
     \begin{subfigure}[b]{0.40\textwidth}
         \centering
         \includegraphics[width=\textwidth]{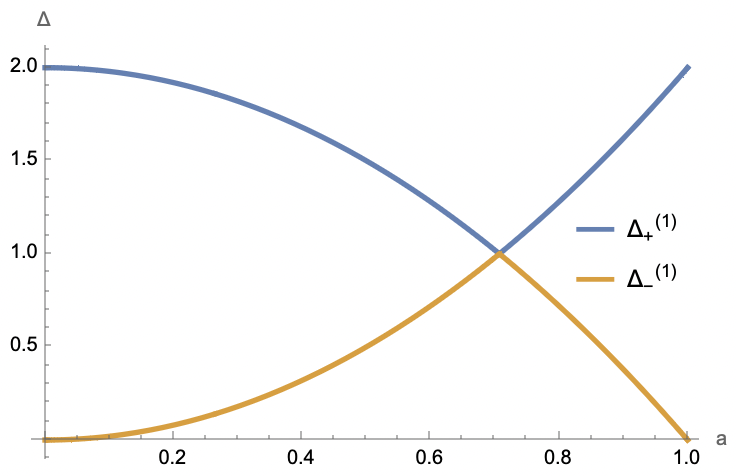}
         \caption{$\Delta_\pm^{(1)}$ for $a<1$}
     \end{subfigure}
     \hfill
     \begin{subfigure}[b]{0.40\textwidth}
         \centering
         \includegraphics[width=\textwidth]{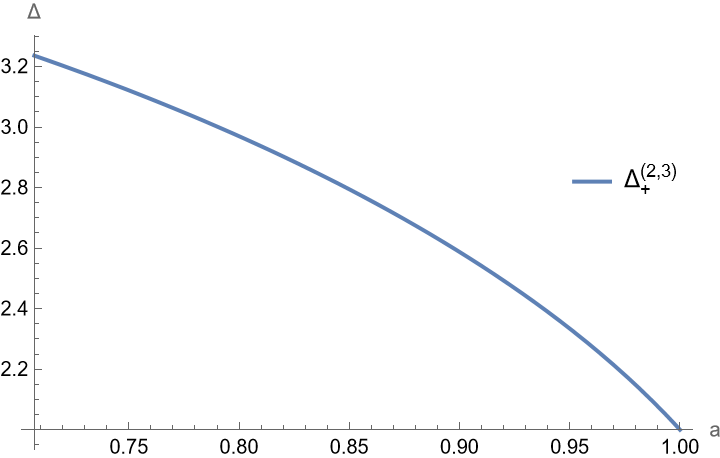}
         \caption{$\Delta_+^{(2,3)}$ for $\frac{1}{ \sqrt{2}}<a<1$}
     \end{subfigure}
        \caption{Conformal dimension of operator dual to fluctuation around extrema.}
        \label{figure:plotdelta}
\end{figure}

For any value of $a$  there is an extremum at $\phi=\phi^{(1)}=0$.  Expanding around it allows to read off the mass  of the  small fluctuation around $\phi=\delta \phi$
\begin{align}
    V\sim -2-2 a^2(1-a^2) \delta \phi^2 + o(\delta \phi^4)
\end{align}
As mentioned before we have $l^{(1)}_{AdS}=1$
Using the standard relation of the mass and conformal dimension of the dual operator one obtains
\begin{align}
    \Delta^{(1)}_\pm =  1 \pm  |1-2 a^2|
\end{align}
which is valid for all $a\in R$.  This implies that the dual operator is relevant for $0<a<1$ and irrelevant for $a>1$.
For $\frac{1}{\sqrt{2}} <a<1$, there are two additional extrema of the potential located at
\begin{align}
    \phi^{(2),(3)}=\pm \frac{1}{ \sqrt{2}a}\ln\left( \frac{1+ 2 a\sqrt{1-a^2}}{ 2a^2-1}\right)
\end{align}
Expanding $\phi = \phi^{(2,3)} + \delta \phi$, gives
\begin{align}\label{ex23}
V= -\frac{2a^4}{ 2a^2-1}- \frac{4a^4(a^2-1)}{ 2a^2-1}\delta \phi^2+ o(\delta \phi^3)
\end{align}
The $AdS_3$ vacuum has a curvature radius 
\begin{align}
    l_{AdS}^{(2,3) }= \frac{\sqrt{2a^2-1}}{ a^2}
\end{align}
and from  (\ref{ex23}) we can read off the mass and determine the conformal dimension of the operator dual to the scalar fluctuation around the extremum.
\begin{align}
    \Delta^{(2,3)}_+ = 1+\sqrt{1+ 8(1-a^2)}
\end{align}
Consequently, the dual operator will always be irrelevant for the values of $a$ where the additional extrema and $AdS$ vacua exist (see figure \ref{figure:plotdelta}).

The simplicity of the minimal  gauged supergravity makes the construction of analytic, as well as numerical solutions, relatively easy.  For example, Poincare sliced domain wall solution representing RG-flows have been constructed in \cite{Deger:2002hv, Arkhipova:2024iem,Golubtsova:2024dad,Golubtsova:2024odp} and string and vortex solutions have been constructed in \cite{Deger:1999st,Abou-Zeid:2001inc,Deger:2004mw,Deger:2006uc,Karndumri:2015sia}. In this note, we utilize an $AdS_2$ slicing ansatz to find Janus and RG-flow interface solutions in this theory.

\section{Janus and RG-interfaces}\label{sec3}

The equations of motion following from the $\theta=A_\mu=0$ truncation of the action  (\ref{actiona}) are

\begin{align}\label{eqofma}
 R_{\mu \nu}-\frac{1}{2} g_{\mu \nu} R&=\frac{1}{2}\left(\partial_\mu \phi \partial_\nu \phi-\frac{1}{2} g_{\mu \nu} \partial_\sigma \phi \partial^\sigma \phi\right)-\frac{1}{2} g_{\mu \nu} V(\phi)\nonumber  \\
0&= \frac{1}{\sqrt{-g}} \partial_\mu\left(\sqrt{-g} g^{\mu \nu} \partial_\nu \phi\right)- V^{\prime}(\phi) 
\end{align}

The ansatz for Janus and RG-flow interfaces is given by taking an $AdS_2$ slicing of the three dimensional metric and demanding that the scalar field $\phi$ only depends on the slicing coordinate $u$.

\begin{align}\label{adsslice}
     ds^2 = du^2+ e^{2 B(u)} \frac{dx^2-dt^2}{ x ^2}, \quad \quad \phi=\phi(u) 
  \end{align}

The equations of motion  (\ref{eqofma}) then become a system of second order ordinary differential equations, for $B$ and $\phi$
\begin{align}
B'' +2 (B')^2 + V + e^{- 2B} & =0 \label{eqom1} \\
{\phi''}+2 B' \phi'  -\frac{\partial V}{ \partial \phi}&=0
\label{eqom2}
\end{align}
Subject to a constraint 
\begin{align}\label{eqofm3}
    (B')^2-\frac{1}{ 4} (\phi')^2+e^{-2 B} +\frac{1}{ 2}V&=0 
\end{align}
In order to determine whether Janus or RG-flow interface solutions exist which preserve some supersymmetry, it is sufficient to, first, consider the vanishing  of gravitino variation in the $AdS_2$ direction 
\begin{align}
    \delta \psi_t &= \partial_t \epsilon +\frac{1}{ 2z} \gamma_0 \Big(-\gamma_1 + B'e^B\gamma_2+ e^B W\Big)\epsilon =0\nonumber \\
     \delta \psi_z &= \partial_z \epsilon +\frac{1}{ 2z} \gamma_1\Big( B'e^B \gamma_2+e^B W\Big)\epsilon =0
\end{align}
where the integrability $(\partial_t \partial_z- \partial_z\partial_t )\epsilon=0$  condition produces the following  equation
\begin{align}\label{susy1}
    1-e^{2B}  W^2+ e^{2B} (B')^2 =0
\end{align}
Secondly,  the dilatino variation
\begin{align}\label{susy3}
    \delta \lambda = -\frac{1}{2} \Big(\gamma_2 \phi' + \frac{2}{ a} \frac{\partial W}{  \partial \phi }\Big) \epsilon
\end{align}
corresponds to a projector on the susy parameter $\epsilon$ if
\begin{align}\label{susy2}
    (\phi')^2 = \frac{4}{ a^2} \left(\frac{\partial W}{\partial \phi}\right)^2
\end{align}
It is straightforward to verify that the conditions (\ref{susy1}) and (\ref{susy2}) are inconsistent with the equations of motion (\ref{eqom2}) unless $\phi=\phi^{(1)}=0$ which is  the supersymmetric  $AdS_3$ vacuum. Consequently, the additional $AdS_3$  $\phi=\phi^{(2,3)}$ which exist for $\frac{1}{ \sqrt{2}} <a<1$ as well as any $AdS_2$ sliced flow solution for which $\phi'$ is not vanishing, will break all the supersymmetries.

It is possible to rewrite the equations of motion  (\ref{eqom1})- (\ref{eqom2}) as a system of first order equations, however as pointed out already in \cite{DeWolfe:1999cp} this is not very useful in obtaining closed form or even numerical solutions.
Here we will employ  the following strategy to obtain numerical solutions of the equations of motion: The $uu$ component of Einstein equations (\ref{eqom2}) is a constraint for reparametrizations of the coordinate $u$. If it is imposed at a fixed $u$  it will be satisfied for all $u$ for solutions of the second order equations of motion. In addition, we look for Janus or RG-interface solutions. These all have the feature  that the warping factor $e^{2B}$ has a minimum. Other solutions are possible but they will generally develop a naked singularity or become non-physical (for example $B$ will diverge or the signature of the metric changes).

Consequently, we impose the initial conditions at the turning point where $B'=0$ which we set by a translation  of the coordinate $u$ to be localized at $u=0$. The constraint (\ref{eqofm3}) then becomes
\begin{align}
 \left. (\phi')^2- 2 V - 4 e^{-2B}\right|_{u=0}=0
\end{align}
and one can determine $B(0)$ from specifying the initial conditions  $\phi'(0)$ and $\phi(0)$.  The numerical solutions can then be obtained by integrating the second order equations (\ref{eqom1}) and (\ref{eqom2}) using a shooting method in Mathematica.
\begin{figure}[ht]
     \centering
     \begin{subfigure}[b]{0.32\textwidth}
         \centering
    \includegraphics[width=\textwidth]{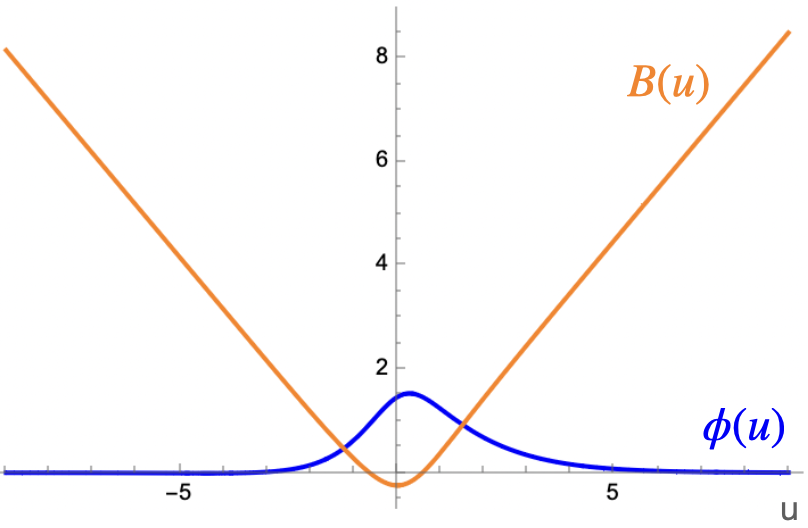}
         \caption{Janus interface with $\phi^{(1)}$ vacuum on both sides}
     \end{subfigure}
     \hfill
     \begin{subfigure}[b]{0.32\textwidth}
         \centering
         \includegraphics[width=\textwidth]{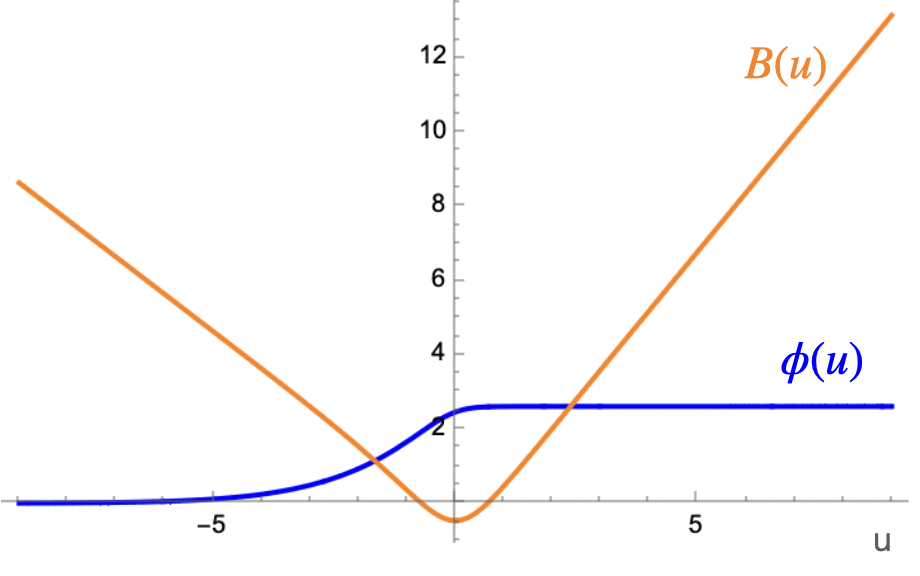}
         \caption{RG-flow interface interpolating between $\phi^{(1)}$ and $\phi^{(2)}$ vacua }
     \end{subfigure}
     \hfill
\begin{subfigure}[b]{0.32\textwidth}
         \centering
         \includegraphics[width=\textwidth]{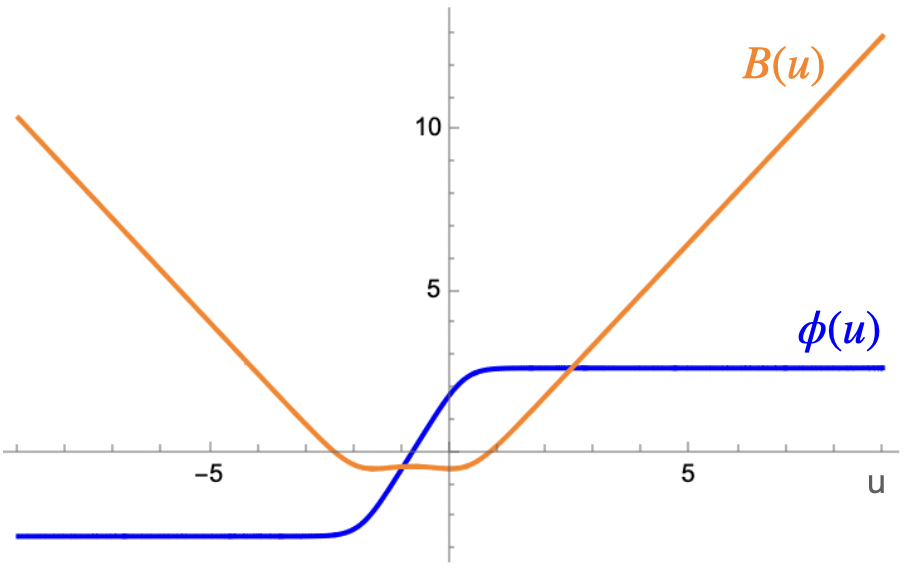}
         \caption{RG-flow interface interpolating between $\phi^{(3)}$ and $\phi^{(2)}$ vacua }
     \end{subfigure}
        \caption{Examples of interface solutions for representative initial conditions }
        \label{figure:plotinter}
\end{figure}

We will illustrate this for the example $a=\frac{3}{ 4}$ for which the potential has  three extrema. There are three types of interface solutions as illustrated for some representative initial conditions in figure  \ref{figure:plotinter}.

\begin{figure}[ht]
     \centering         \includegraphics[width=.45\textwidth]{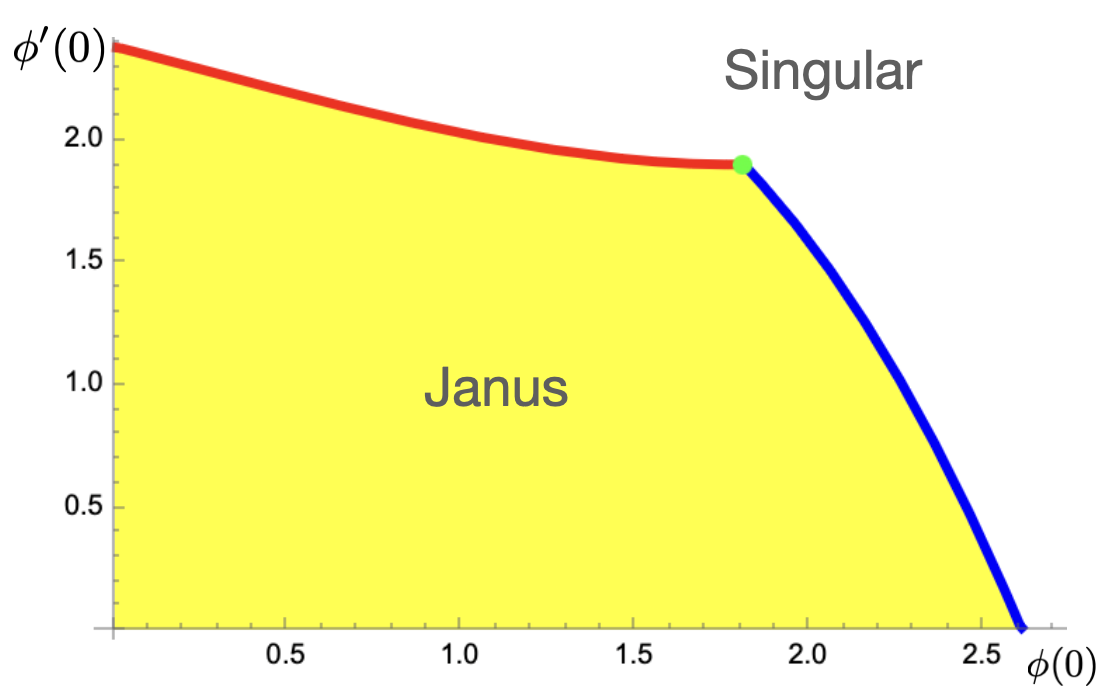}
         \caption{Phase diagram for interface solutions for the $a=0.75$. The diagram is extended to the other quadrants using  $\phi(0)\to -\phi(0)$ and $\phi'(0)\to -\phi'(0)$ maps.}
         \label{phasedia}
\end{figure}

The plot (a) depicts a Janus interface between the supersymmetric vacuum $\phi^{(1)}$ as $u\to \pm \infty$. To obtain such a solution the initial conditions do not have to be fine-tuned, in figure \ref{phasedia} the  initial conditions leading to Janus solutions  are in the yellow area.
The plot (b) depicts an RG-flow interface interpolating between the supersymmetric vacuum $\phi^{(1)}$ as $u\to -\infty$ and the vacuum $\phi^{(2)}$ as $u\to \infty$. The initial conditions have to be fine-tuned in figure \ref{phasedia} where they correspond to the blue line. The red line in figure \ref{phasedia} corresponds to initial conditions which lead to RG-flow interface interpolating between the supersymmetric vacuum $\phi^{(3)}$ as $u\to -\infty$ and the vacuum $\phi^{(1)}$ as $u\to \infty$.   The plot (c) corresponds to a solution that interpolates between the vacuum $\phi^{(3)}$ as $u\to -\infty$ and the vacuum $\phi^{(2)}$ as $u\to \infty$. Here both initial conditions have to be fine-tuned and in figure \ref{phasedia} this solution corresponds to the green dot. Initial conditions outside the colored region lead to solutions which develop a naked singularity at a finite value of $u$.

For values of $a< \frac{1}{ \sqrt{2}}$ only the supersymmetric vacuum $\phi^{(1)}$ exists and the solutions are all Janus solutions which look qualitatively similar to (a) in  figure \ref{figure:plotinter}. For value  $a>1$ all interface solutions develop naked singularities.

\section{Holographic observables}
\label{sec4}

The numerical solutions obtained in the previous sections can be used to calculate holographic observables.   Here we focus on a few, namely  the entanglement entropy of an interval both symmetrically about the interface \cite{Azeyanagi:2007qj,Chiodaroli:2010ur} and at the interface \cite{Gutperle:2015hcv,Karch:2021qhd}, as well as  the transmission coefficient \cite{Bachas:2020yxv,Bachas:2022etu, Baig:2023ahz,Baig:2024hfc}.

\begin{figure}[ht]
     \centering
     \begin{subfigure}[b]{0.45\textwidth}
         \centering
         \includegraphics[width=\textwidth]{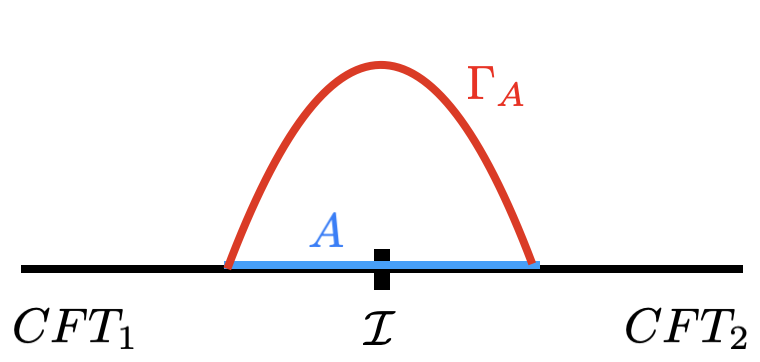}
         \caption{Entangling surface $A$ symmetric about interface $I$ }
     \end{subfigure}
     \hfill
     \begin{subfigure}[b]{0.45\textwidth}
         \centering
         \includegraphics[width=\textwidth]{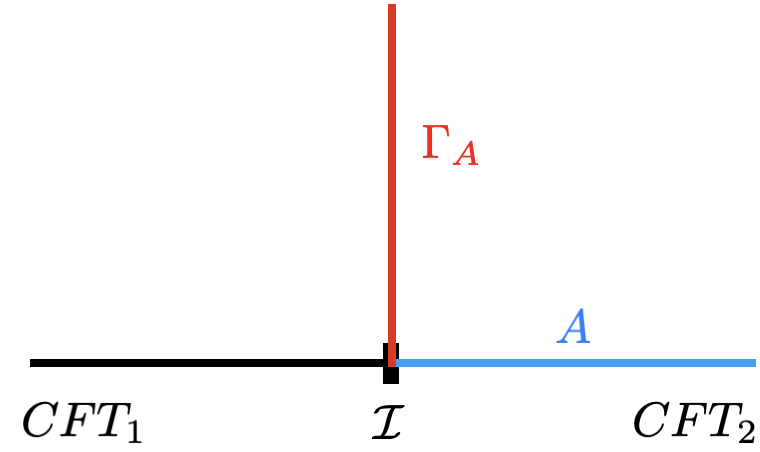}
         \caption{Entangling surface $A$ at interface $I$ }
     \end{subfigure}
        \caption{}
        \label{figure:interface}
\end{figure}

 Correlations functions in the background of Janus solutions  have been discussed in  \cite{Papadimitriou:2004rz,Chiodaroli:2016jod,Melby-Thompson:2017aip,Ghodsi:2022umc}.  However  these correlators  are  more difficult to obtain if the RG-flow  solutions of the background are known only numerically. There are other holographic observables such as the on-shell action and volume measures of complexity (see e.g. \cite{Auzzi:2021nrj,Auzzi:2021ozb,Bak:2015jxd}) which will not be discussed in this note.

\subsection{Symmetric entanglement entropy}\label{syment}
The Ryu-Takayanagi prescription \cite{Ryu:2006bv,Ryu:2006ef} allows to calculate the entanglement entropy holographically, for an interval $A$ the entanglement entropy is given by
\begin{align}\label{RTformula}
    S_{EE}(A) =\frac{ {\rm Length}[\Gamma_A]}{  4 G_N}
\end{align}
where $\Gamma_A$ is the geodesic in the bulk of spacetime which ends at the interval $A$ on the boundary. For  $AdS_2$ sliced  solution describing an interface, there are two simple geometries one can consider. Firstly, we can choose the interval to be symmetric about the interface \cite{Azeyanagi:2007qj,Chiodaroli:2010ur, Karch:2021qhd,Afxonidis:2024gne}  and the entanglement entropy takes the following form 
\begin{align}\label{sares}
    S_A = \frac{c_L+ c_R}{ 6}\ln \frac{l}{ \epsilon} + \ln g_A
\end{align}
Here $c_{L/R}$ are the central charges of the CFTs on either side of the interface. For a Janus interface they are equal, whereas for an RG-flow interface, they will be different. 
Furthermore, $2l$ is the length of the interval $A$, which is symmetric about the interface and  $\epsilon$ is a UV cutoff and $\ln(g_A)$ is the g-factor (or interface entropy) which is  a physical quantity associated with the number of degrees of freedom localized on the interface.

To apply the Ryu-Takayanagi formula (\ref{RTformula}) for the $AdS_2$ sliced metric (\ref{adsslice}),  it was shown in \cite{Chiodaroli:2010ur} that the geodesic is parameterized  by  choosing a fixed   $z=l$ and $u\in [-\infty, \infty]$.  This  geodesic corresponds to an  entanglement interval symmetric about the interface at the origin, i.e.  $A=[-l,l]$.   In the following, we apply  the holographic calculation \cite{Chiodaroli:2010ur,Karch:2021qhd,Afxonidis:2024gne}, where details can be found. 
The length of the geodesic is divergent 
\begin{align}\label{geodes}
 {\rm Length}[\Gamma_A]=\int_{u_{-\infty}}^{u_{\infty}} du=u_{\infty}-u_{-\infty} 
\end{align}
and must  be regulated by mapping the $AdS_2$ sliced metric (\ref{adsslice}) in the asymptotic regions $u\to \pm \infty$  to a Fefferman-Graham coordinate and  then  introducing a uniform UV cutoff $\epsilon$.
For the RG-flow and Janus interface solutions obtained in section \ref{sec3} the warp factor takes the following form for large $|u|$
\begin{align}
    \lim_{u\to +\infty}B(u) &= \frac{u}{ L_R} + \ln \gamma_R + o(\frac{1}{ u}), \quad \quad  \lim_{u\to -\infty}B(u) = -\frac{u}{ L_L} - \ln \gamma_L + o(\frac{1}{ u}) 
\end{align}

For large $u\to \pm \infty $  the $AdS_2$ sliced metric can be mapped asymptotically to a Poincare sliced $AdS_3$ with radius $L_{R/L}$ respectively by the following coordinate change
 \begin{align}\label{largeufit}
u\to +\infty: \quad      u = L_R \Big(\ln \frac{x}{ z} -\ln \gamma_R+ \ln L_R\Big) + o(z),\nonumber\\
u\to -\infty: \quad      u = -L_L \Big(\ln \frac{x}{ z} -\ln \gamma_L+ \ln L_L\Big) + o(z)
 \end{align}
Here $z$ is the radial coordinate in the asymptotically $AdS_3$ in Poincare coordinates and the Fefferman-Graham UV-cutoff is $z=\epsilon$. The boundary  of the entangling surface is located at $x=l$. Using The Brown-Henneaux formula for the central charge on the left and right sides of the interface
\begin{align}
   \frac{c_{L/R}}{ 6} = \frac{ L_{L/R}}{ 4 G_N} 
\end{align} it follows that the holographic entanglement entropy (\ref{RTformula}) takes the form (\ref{sares}) with the g-factor given by
\begin{align}
    \ln g_A = -\frac{1}{ 6 } \Big( c_R \ln \frac{\gamma_R}{ L_R}+ c_L \ln \frac{\gamma_L}{ L_L}\Big)
\end{align}
For the numerical Janus or RG-flow interfaces obtained in section \ref{sec3} this can be calculated by fitting $B(u)$ for large $|u|$ to (\ref{largeufit}) to obtain $L_{L/R}$ and $\ln(\gamma_{L/R})$.   We illustrate this here by presenting a plot of $\ln g_A$ for the RG-flow  interfaces interpolating between two distinct vacua  for initial conditions given   by the red and blue curves in   figure  \ref{phasedia}.
\begin{figure}[ht]
     \centering         \includegraphics[width=.50\textwidth]{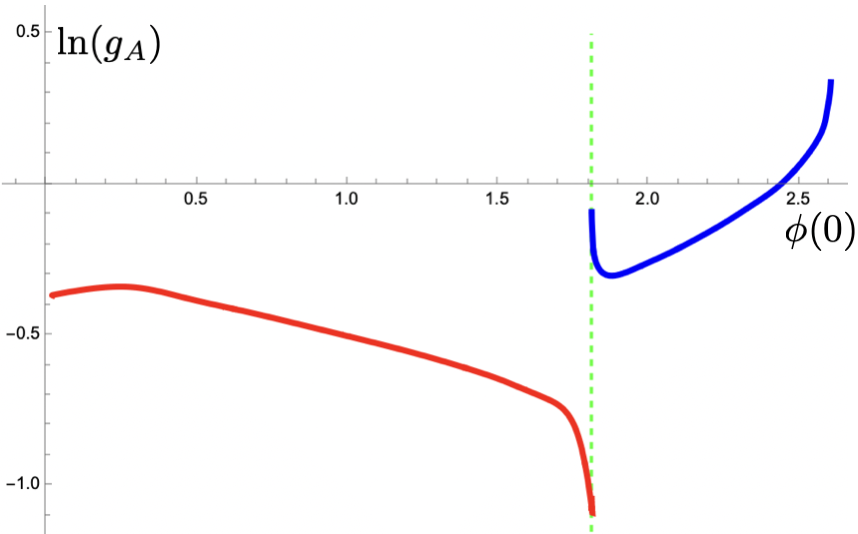}
         \caption{Interface entropy $\ln(g_A)$ for the RG-flow interfaces depending on the initial condition $\phi(0)$.}
         \label{lnga}
\end{figure}

\subsection{Entanglement entropy at the interface}
Secondly, one can consider an entanglement interval $A$  which ends on the defect. For simple CFTs the entanglement entropy can be calculated using the replica trick \cite{Sakai:2008tt,Brehm:2015lja} and it takes the following form
\begin{align}\label{ceffa}
    S_A&=\frac{c_{eff}}{ 6} \ln \frac{l}{ \epsilon}
\end{align}
Where $c_{eff}$ is an effective central charge, which depends on the details of the interface  and measures the amount of entanglement across the interface.

The entanglement entropy at the interface has been calculated holographically in
\cite{Gutperle:2015hcv} where it has been shown for the AdS sliced metric (\ref{adsslice})   that the Ryu-Takanayagi geodesic is along the $z$ coordinate and  $u$ is fixed at the minimum of $B(u)$, which was chosen to be at $u=0$ for the numerical solutions constructed in section \ref{sec3}. One obtains for the numerical solutions of section \ref{sec3}
\begin{align}\label{atinterfacea}
    S_{A} =\frac{l}{ 4 G_N } e^{B(0)} \int \frac{dz}{ z} = \frac{1}{ 4 G_N}e^{B(0)} \log\frac{l}{ \epsilon} 
\end{align}
Here $\epsilon$ is a UV cutoff and $l$ is the length of the interval which we take to to be very large in order to eliminate the contribution from the other end of the interval.
Hence the effective central charge (\ref{ceffa}) is given by
\begin{align}
   c_{eff} = \frac{3}{ 2 G_N } e^{B(0)} 
\end{align}

The effective central charge $c_{eff}$ has interesting properties such as a universal bound  \cite{Karch:2024udk} and a relation to the transmission coefficient discussed below for certain supersymmetric Janus solutions  \cite{Baig:2024hfc}.

\subsection{Transmission and reflection coefficients}\label{sec4.3}

Another quantity is the energy reflection and transmission amplitude  which describe the flow though and reflection of energy from the CFT interface.  In the CFT the transmission amplitude can be expressed as a normalized two-point function of the stress tensors $T_{1,2}$ of the CFTs on the two sides of the interface \cite{Bachas:2001vj,Quella:2006de,Meineri:2019ycm}
\begin{align}
    {\cal T} = \frac{c_{LR} }{ c_L+c_R} =\frac{ \langle T_1 T_2 + \bar T_1 \bar T_2\rangle}{  \langle (T_1+\bar T_1) (T_2+ \bar T_2)\rangle}
\end{align}
where $c_L$ and $c_R$ are the central charges of the two CFTs on either side of the interface. The reflection amplitude ${\cal R}$ is determined by unitarity ${\cal R}+{\cal T}=1$.
A holographic expression for $c_{LR}$ has been obtained in \cite{Bachas:2022etu}, by taking  a continuum limit for the reflection and transmission of energy in   an array of probe branes. 
\begin{align}\label{clrresult}
    c_{LR} =\frac{3}{G_N}\left(\frac{1}{ l_R}+\frac{1}{ l_L}+ 8 \pi G_N \sigma\right)^{-1}
\end{align}
Where $l_{L,R}$ are the AdS radius of the asymptotic $AdS_3$ half-regions close to the boundary on either side of the interface.  The quantity $\sigma$  depends on the scalar field kinetic energy, for the action (\ref{actiona}) is given by
\begin{align}
\sigma = \int_{-\infty}^{\infty}(\phi')^2 du
\end{align}

\begin{figure}[ht]
     \centering         \includegraphics[width=.46\textwidth]{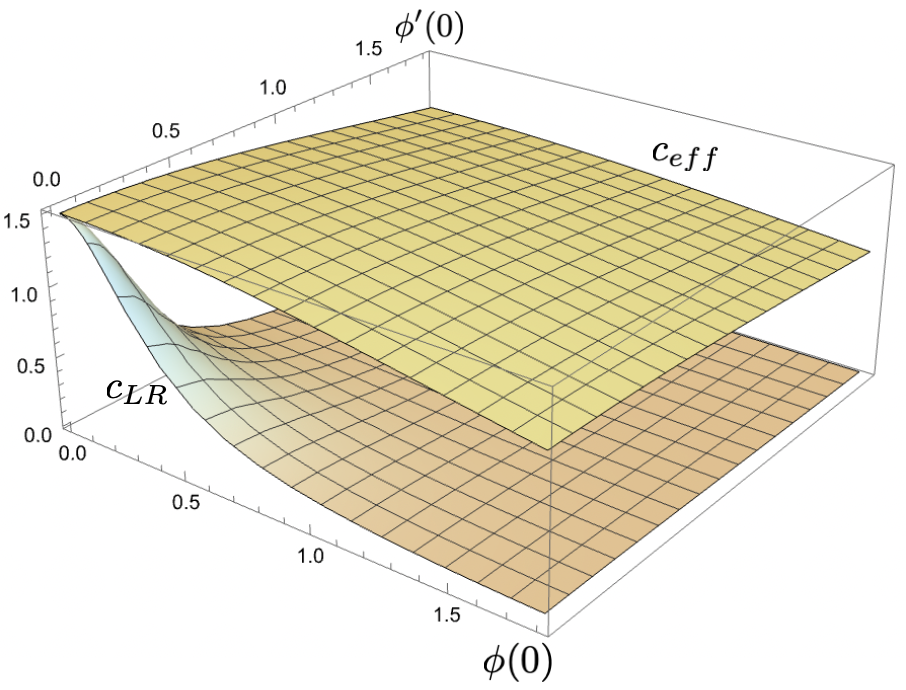}
         \caption{Plot of $c_{LR}$ and $c_{eff}$ for $a=\frac{3}{ 4}$ of a function of initial conditions $\phi(0), \phi'(0)$.}
         \label{inequa}
\end{figure}

and can be calculated using the numerical solution obtained in the previous section, for both the Janus solution as well as the RG-flow interface solution.

In \cite{Karch:2024udk} a set of inequalities relating $c_{LR}$ (and hence the transmission coefficient ${\cal T}$) to $c_{eff}$ and the central charges on either side of the interface was proposed
\begin{align}
    0\leq c_{LR} \leq c_{eff}\leq {\rm min}(c_L,c_R)
\end{align}

In \cite{Karch:2024udk} some holographic and CFT examples were checked and it was argued that the inequality between $c_{LR}$ and $c_{eff}$ is only becoming an equality for a completely reflective or transmissive interface. We can use the numerical solutions in our simple supergravity model to check and we found that the strict inequality holds for all Janus and RG-flow solutions. We can illustrate the validity of this inequality  with the plot in figure \ref{inequa} of $c_{LR}$ and $c_{eff}$ for  $a=\frac{3}{ 4}$. Note that the point in the plot where $c_{LR}=c_{eff}$ corresponds to $\phi(0)=\phi'(0)$ which is the supersymmetric $AdS_3$ vacuum and hence corresponds to a trivial topological interface , i.e. no interface at all.

\section{Transmission coefficient for  $N=8,d=3$ gauged supergravity}
\label{sec5}

In the previous section  entanglement entropy and reflection coefficients  were calculated for the non-supersymmetric Janus and RG-flow interfaces in minimal $N=2$ gauged supergravity. Recently, it has been observed \cite{Baig:2024hfc} that there is a relation of the transmission coefficient and the entanglement entropy at the interface for a class of supersymmetric Janus solutions constructed as $AdS_2\times S_2\times T_4\times \Sigma_2$ solutions of type IIB supergravity in \cite{Baig:2024hfc}. In this section we use supersymmetric Janus solutions of $d=3,N=8$ gauged supergravity which were obtained some time ago \cite{Chen:2020efh} to show that the relation of these two quantities holds for these solutions as well.

  In the following, we will follow the construction \cite{Nicolai:2001ac}. 
The scalar fields of $d=3$, $N=8$ gauged supergravity take values in  a $G/H = SO(8, n)/\big(SO(8) \times SO(n))$ coset. There are   $8n$ independent scalar degrees of freedom.
The three dimensional theory can be constructed as  a truncation of six-dimensional $N = (2, 0)$ supergravity on $AdS_3 \times S^3$ coupled to $n_T \geq 1 $ tensor multiplets, where the number of tensor multiplets is fixed by $n_T = n-3$.
The special  cases $n_T = 5$ and $21$ are related  to compactifications of ten-dimensional type IIB on $T^3$ and $K3$, respectively and hence are related to low energy limits of consistent string theories. Smaller values of $n$ can  be obtained by consistent truncations,  see \cite{Samtleben:2019zrh} for a discussion of  consistent truncations of six-dimensional $N = (1, 1)$ and $N = (2, 0)$ using exceptional field theory.     The action, gauging and supersymmerty transformations were constructed in \cite{Nicolai:2001ac} using the embedding tensor formalism. The details of the action and the construction of the half-BPS Janus solution can be found  in  \cite{Chen:2020efh}

The Janus solution  considers the simplest case with $n=1$ for which  there are eight coset scalars $\phi_i, i=1,2,\cdots 8$. It was shown in \cite{Chen:2020efh} that one can further consistently truncate the theory   where only  two denoted as $\phi_4,\phi_5$ have  a nontrivial profile and all others are set to zero.  The truncated bosonic action takes the following form
\begin{align}\label{n8action}
    S&= \frac{1}{2} \int d^3 x \sqrt{-g}\Big\{ R-    P_\mu^I P^{\mu\; I }-V \Big\}
\end{align}
Where the notation  $\Phi= \sqrt{\phi_4^2+\phi_5^2}
$ is used for compactness.  The kinetic energy term and the potential for non-vanishing scalars are given by
\begin{align}
    P_\mu^I P^{\mu\; I }&= \frac{\phi_4^2+(\sinh^2 \Phi +\phi_4^2) \phi_5^2}{ \Phi^4} \partial_\mu \phi_4\partial^\mu \phi_4-\frac{\phi_5^4+(\sinh^2 \Phi +\phi_5^2) \phi_4^2}{ \Phi^4} \partial_\mu \phi_5\partial^\mu \phi_5 \nonumber \\
   & \quad- \frac{2 (\Phi^2-\sinh^2 \Phi) \phi_4\phi_5}{ \Phi^4} \partial_\mu \phi_4\partial^\mu \phi_5\nonumber\\
   V&=-\Big( \frac{\sinh^2(\Phi) \phi_4^2} { \Phi^2}+2\Big)
\end{align}

It was shown in \cite{Chen:2020efh} that a solution of the equations of motion which  preserves  half the supersymmetries of the $N=8$ gauged supergravity is given by the scalar profiles  $\phi_4(u), \phi_5(u)$ which are implicitly  defined  and depend on two real parameters $p,q$
	\begin{align} 
	\frac{|\phi_4| \sinh \Phi}{\Phi} &= |\sinh q|\;  {\rm sech} u \nonumber \\
	\frac{\phi_5 \sinh \Phi}{\Phi} &=  \sinh p \cosh q + \cosh p \sinh q \tanh u \label{eq:scalar-sol}
	\end{align}
 The $AdS_2$ sliced metric is given by
 \begin{align}\label{n8metric}
     ds^2 = du^2 + {\rm sech^2 q}\cosh^2 u \frac{d\xi ^2-dt^2 }{ \xi^2}
 \end{align}

\begin{figure}[ht]
     \centering
     \begin{subfigure}[b]{0.45\textwidth}
         \centering
         \includegraphics[width=\textwidth]{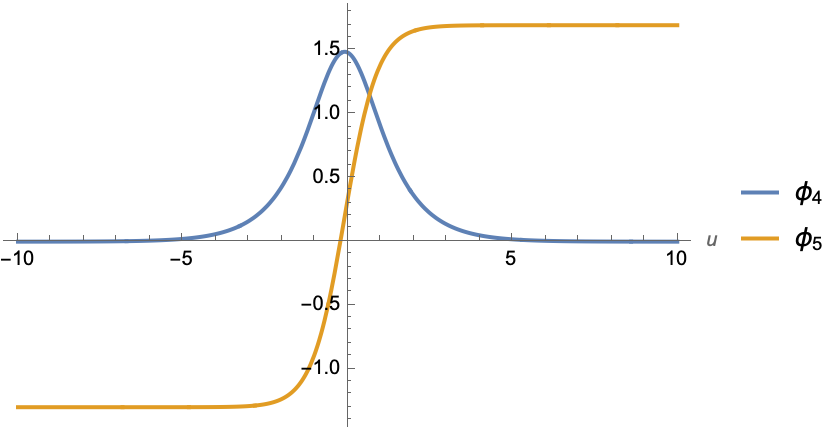}
         \caption{Scalar profile for $\phi_4, \phi_5$ for $q=\frac{3}{ 2}, q=\frac{1}{ 5}$}
         \label{N25a}
     \end{subfigure}
     \hfill
     \begin{subfigure}[b]{0.45\textwidth}
         \centering
         \includegraphics[width=\textwidth]{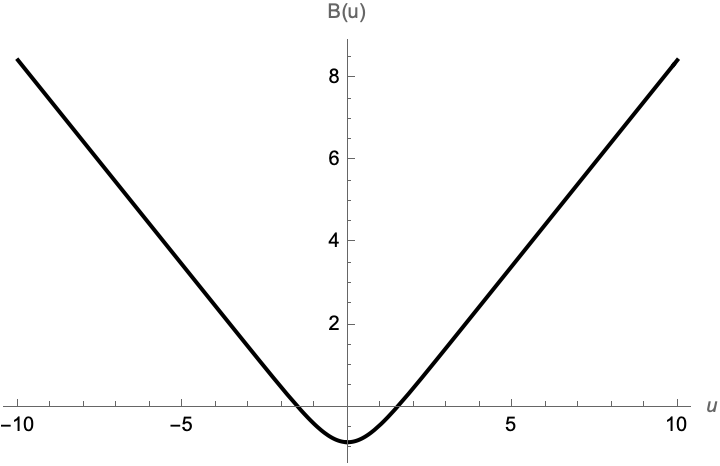}
         \caption{Warp factor $B$ for $q=\frac{3}{ 2}, q=\frac{1}{ 5}$}
         \label{N25b}
     \end{subfigure}
        \caption{Scalar Profiles and warp factors for the half-BPS solution.}
        \label{figure:n=8}
\end{figure}

 A  plot of the scalars and warp factor as a function of the slicing coordinate $u$ is presented in figure \ref{figure:n=8}.
 Note that the solution with $q=0$ corresponds to the unit radius $AdS_3$ vacuum where the massless scalar is constant and given by $\phi_5=\sinh p$.   The $g$ factor for a symmetric entanglement entropy for this  solution was calculated 
in \cite{Chen:2020efh} and can easily be reproduced using the expression given in section \ref{syment} and one obtains
\begin{align}
   \ln( g_A) = \frac{c}{3} \ln(\cosh q)
\end{align}
For the entanglement interval at the interface, we can use (\ref{atinterfacea}) and the metric (\ref{n8metric}) to obtain the holographic result for the effective central charge
\begin{align}
    \frac{c_{eff}}{ c} = e^{B(0)}= \frac{1}{ \cosh q}
\end{align}
For the transmission coefficient, the relevant quantity is the integral of $d\sigma $ which for the action (\ref{n8action}) is given by (choosing units such that $8\pi G_N=1$).
\begin{align}
    \sigma = \int_{-\infty }^\infty du \; P_u^I  P_u^I = 2 \sinh^2 q
\end{align}
Plugging this result in the expression (\ref{clrresult}) and noting that $l_R=l_L=1$ one obtains 
\begin{align}
    {c_{LR}}  =  2c \frac{1}{ 2(1+ \sinh^2 q)} = \frac{c}{\cosh^2q}
\end{align}
Hence we observe that the transmission coefficient and the effective central charge obey the following relation
\begin{align}
    \frac{c_{LR}}{c}= \left(\frac{c_{eff}}{ c}\right)^2
\end{align}
The same relation was found in \cite{Baig:2024hfc} for the ten dimensional half BPS Janus Janus solution of \cite{Chiodaroli:2010ur,Chiodaroli:2009yw}. Note that the exact relation has a different form than the inequalities discussed at the end of section \ref{sec4.3}, it is however easy to verify that the supersymmetric solutions also satisfy these inequalities.

\section{Discussion}
\label{sec6}
In this paper, we used  minimal $d=3,N=2$ gauge supergravity with a single scalar field to construct solutions that represent Janus and RG-flow interface solutions. The model depends on a single parameter $a$, for $a< \frac{1}{ \sqrt{2}}$ there is a single supersymmetric $AdS_3$ vacuum and $AdS_2$ sliced solution are Janus interface solution. For  $ \frac{1}{ \sqrt{2}}<a<1$ there are two additional non-supersymmetric vacua and depending on initial conditions there are Janus solutions as well as fine-tuned RG-flow interface solutions that interpolate between the different vacua. We showed that the single scalar model does not allow for interface solutions that preserve any supersymmetry and solutions are obtained by numerical integration.  We calculated holographic observables such as symmetric and interface entanglement entropy and transmission coefficients using the numerical solutions and confirmed that the inequalities involving $c_{LR}$ and $c_{eff}$ proposed in \cite{Karch:2024udk} are satisfied for the solutions obtained in this paper.  The simplicity of the model and solutions makes it a good model to calculate other holographic observables such as correlation functions, complexity measures or other entanglement entropy and check whether other inequalities involving these quantities can be discovered.

For supersymmetric Janus solutions previously obtained in \cite{Chen:2020efh} we showed that an exact relation between the entanglement entropy and the reflection coefficient first obtained in \cite{Baig:2024hfc} is satisfied. It would be interesting to find a proof of this relation for all supersymmetric $AdS_3$ Janus solutions, but we have not been able to find one so far.
We leave these interesting questions for future work

\acknowledgments
The work of M.G. was supported, in part, by the National Science Foundation under grant PHY-2209700. 
The authors  are grateful to the Bhaumik Institute for support. We thank the authors of \cite{Baig:2024hfc} for an interesting correspondence regarding their paper.

\newpage

\providecommand{\href}[2]{#2}\begingroup\raggedright\endgroup

\end{document}